# Letter to the Editor: Note on published research on the effects of COVID-19 on the environment without sufficient depth of science

Ahmed Mustafa, Timon McPhearson
Urban Systems Lab, The New School, NY, USA
a.mustafa@newschool.edu, timon.mcphearson@newschool.edu

Dear Editor-in-Chief:

We have given two articles published recently in Science of the Total Environment by Mandal and Pal (2020) and Zambrano-Monserrate et al. (2020) a thorough reading. Both articles present a significant association between the novel Coronavirus (COVID-19) social distancing policies and improvement in environmental quality such as air pollution, land surface temperature, and noise. Both articles present good research, complemented by detailed explanations and displays, yet we have a few concerns that affect the interpretation and meaning of the results.

First, Mandal and Pal (2020) extracted Particulate Matter (PM), and land surface temperature from the Landsat 8 OLI images for the dates of 12-March-2020 as a representative of pre-lockdown time and 28-March-2020 and 13-April-2020 as the representative of ongoing lockdown periods. However, the authors did not provide detailed information on weather variables during those days which should have a direct impact on both land surface temperature and PM levels. Moreover, the cloud cover of the three Landsat 8 OLI images should be reported since cloud cover is well known to impact land surface temperature and can provide biased results (Lu et al., 2011) .

Mandal and Pal (2020) and Zambrano-Monserrate et al. (2020) reported a dramatic reduction in the concentrations of Nitrogen Dioxide (NO2) and PM with a diameter < 2.5 µm and <10 µm in some parts of the world, Zambrano-Monserrate et al. (2020) Figure 1, and Mandal and Pal (2020) Figures 2 and 3. However, there was not significant incorporation or discussion of the multiple factors that have an interdependent impact on how, and where, different levels of pollution are found. This omission makes it difficult to demonstrate an explicit relationship between decreased air pollution and the stay-at-home orders. Both articles did not analyze or discuss such factors, and believe have led to overstating and misleading scientific conclusions. For example, authors did not report that NO2 and some other air pollutant concentrations tend to be lower in the spring and summer than in the fall and winter so it is normal to see air pollutant levels in April lower than in January, February, or March. This is in line with other studies that have detected similar seasonal patterns in Cabauw (The Netherlands) and Calcutta (India) (Demuzere et al., 2009; Mondal et al., 2000). During cold seasons, atmospheric stability, as a result of frequent inversion layer that happens when the upper air layer is warmer than a lower one, leads to the accumulation of pollutants (Tiwari et al., 2015). Even comparing the same period of 2020 with 2019, as in Zambrano-Monserrate et al. (2020) Figure 2, we should explore air quality regulations and historical trends of air pollutant concentrations in a certain study area. For instance, the annual drop in concentrations of air pollutants including oxides of Nitrogen (NOx) (Figure 1) in the USA (one of the top countries that emit air pollutants), that is largely driven by federal and state implementation of air quality regulations (Sullivan et al., 2018), can easily confuse the relation between potentially cleaner air and the COVID-19.

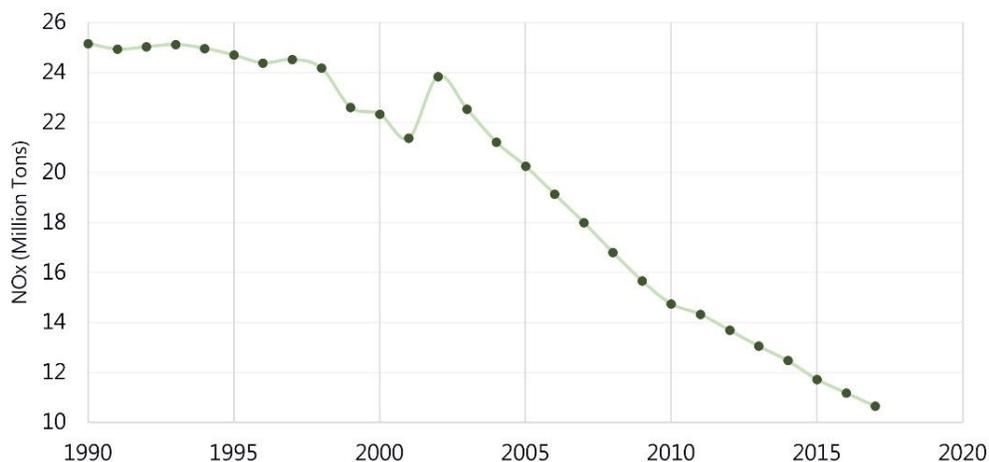

**Figure 1:** $NO_x$ emission trend in USA since 1990.
Data collected from US Environmental Protection Agency.

More importantly, variations in weather conditions are substantial determinants in NO2 and many other air pollutant concentrations (Borge et al., 2019). For example, high wind speed causes the dispersal and dilution of pollutants. Wind can also blow NO2 from areas that have higher NO2 concentrations, e.g., industrial areas, to residential areas causing increased NO2 levels. Precipitation washes out the air and can relatively reduce pollutants in the air whereas air temperatures play an important role in the chemical reactions of pollutants in the air.

Here, we provide a more robust examination of the multiple drivers of air pollution in any given period and plot the average weekly wind speed, temperature, and precipitation in New York City between February 1 and May 1, 2020, to illustrate examples of the complexity of interactions between NO2 and weather conditions (Figure 2). The COVID-19 orders to stay at home and practice social distancing began in New York City on March 16, 2020. An inverse relationship with wind speed with the concentration of NO2 during this time period can easily be observed and can likely be explained by higher wind speed affecting the dispersal of the NO2 concentration. Precipitation and temperature data also reveal inverse relationships with NO2, especially during windy weather conditions. Figure 2 highlights two peaks of NO2 concentration during the weeks started on February 22 (before the lockdown) and April 4 (during the lockdown). Although the peak of April 4 is lower than February 22, the weather variables were not the same. While the wind speed and precipitation rates were slightly different during the two weeks, the temperature was significantly higher in the week of April 4. Higher temperatures help accelerate the oxidization of NO2 in the air and, therefore, lower the NO2 concentration (Khoder, 2002). Additionally, chemical interactions between different air pollutants including NO2, Particulate Matter 2.5 microns, Sulfur Dioxide, and Carbon Monoxide would help to understand the likely multiple drivers of observations of decreased urban air pollution during the COVID-19 pandemic.

Data presented in Figure 2 were collected from the Copernicus Sentinel-5P satellite data which provides one of the most accurate satellite measurements of NO2. To avoid under/overestimation of NO2, we selected only those NO2 retrievals for the clear sky for which cloud radiance fraction ≤ 10%. This is an important procedure as clouds can bias measurements by obscuring the sensor's view of the lower atmosphere affecting both air pollution, and land surface temperature. Thus, we demonstrate how these other rapidly published studies could be, and should be improved. Weekly weather data shown here were collected from Phase 2 of the North American Land Data Assimilation System (NLDAS-2). NLDAS-2, which provides hourly weather data, and is a collaboration project among several groups in North America including NASA, National Oceanic and Atmospheric Administration (NOAA) Princeton University, and the University of Washington.

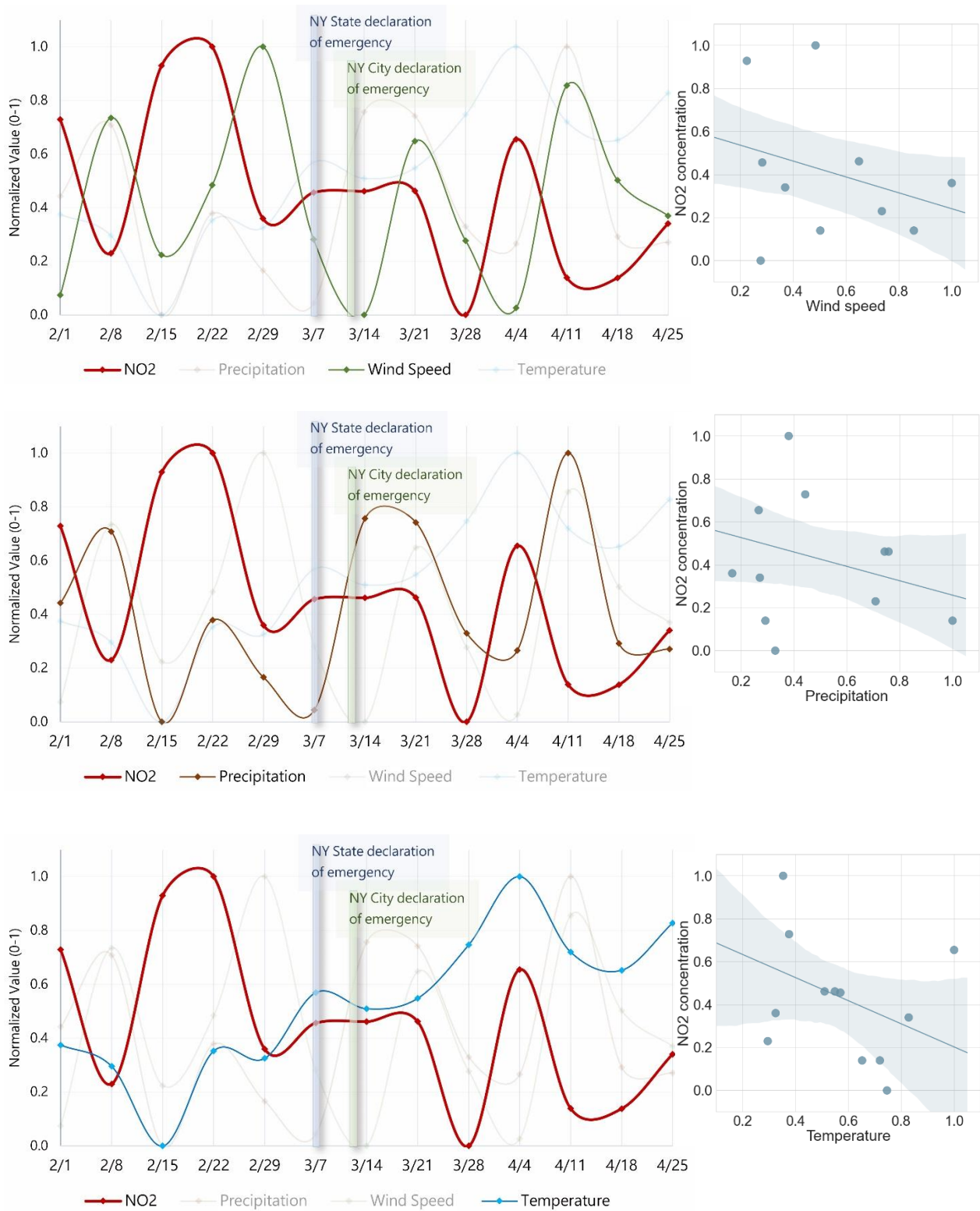

**Figure 2:** (Left) Weekly average NO2, wind speed, precipitation and temperature for week starting from listed date for New York City 5 boroughs between Feb. 1 and May 1, 2020, and (right) linear regression plots and a 95% confidence interval (shaded). All values are normalized between 0 and 1.
Data sources: Sentinel-5P and NLDAS-2.

From our analysis and review of the literature, we find that it remains unclear to what extent social distancing policies and stay-at-home orders that impact human behavior have contributed to observations of improvement of environmental quality such as observed decreases in air pollution. Any study that claims that COVID-19 social distancing policies or similar efforts to reduce the spread of the Coronavirus had a significant impact on the environment must include more comprehensive and detailed research into multiple factors, including at minimum weather variables and air quality regulations, that may be as, or more, important to social distancing policies in explaining temporal changes in environmental quality.

**Funding**: This material was supported by a US National Science Foundation RAPID project, "Interdependent social vulnerability of COVID-19 and weather-related hazards in New York City" (Award #2029918).

6